\documentclass{emulateapj}
\usepackage{apjfonts}

\newcommand{\pivec}{\mbox{\boldmath $\pi$}}
\newcommand{\gammavec}{\mbox{\boldmath $\gamma$}}

\lefthead{UDALSKI ET AL.}
\righthead{ }

\begin{document}
\title{A VENUS-MASS PLANET ORBITING A BROWN DWARF: 
MISSING LINK BETWEEN PLANETS AND MOONS}

\author{
A.~Udalski$^{1}$, 
Y.~K.~Jung$^{2}$, 
C.~Han$^{2,\ast}$, 
A.~Gould$^{3}$,
% OGLE member ----------------
S.~Koz{\l}owski$^{1}$, 
J.~Skowron$^{1}$, 
R.~Poleski$^{1,3}$, 
I.~Soszy{{\'n}}ski$^{1}$, \\
P. Pietrukowicz$^{1}$, 
P. Mr{\'o}z$^{1}$,
M.~K.~Szyma{\'n}ski$^{1}$,
{\L}.~Wyrzykowski$^{1}$, 
K.~Ulaczyk$^{1}$,
G. Pietrzy{\'n}ski$^{1}$,
% --------------------------
Y.~Shvartzvald$^{4}$, \\
D.~Maoz$^{4}$, 
S.~Kaspi$^{4}$, 
B.~S.~Gaudi$^{3}$, 
K.-H.~Hwang$^{2}$, 
J.-Y.~Choi$^{2}$,
I.-G.~Shin$^{2}$, 
H.~Park$^{2}$, 
V.~Bozza$^{5,6}$\\
}

\bigskip\bigskip
\affil{$^{1}$Warsaw University Observatory, Al.~Ujazdowskie~4, 00-478~Warszawa, Poland}  
\affil{$^{2}$Department of Physics, Chungbuk National University, Cheongju 371-763, Republic of Korea} 
\affil{$^{3}$Department of Astronomy, Ohio State University, 140 W. 18th Ave., Columbus, OH  43210, USA} 
\affil{$^{4}$School of Physics and Astronomy, Tel-Aviv University, Tel-Aviv 69978, Israel} 
\affil{$^{5}$Dipartimento di Fisica "E.R. Caianiello", Universit\`a di Salerno, Via Giovanni Paolo II 132, 84084, Fisciano (SA), Italy}
\affil{$^{6}$Istituto Nazionale di Fisica Nucleare, Sezione di Napoli, Via Cinthia 9, 80126 Napoli, Italy}
\affil{$^{\ast}$Corresponding author}

\bigskip\bigskip
%\affil{$^{O1}$Warsaw University Observatory, Al. Ujazdowskie 4, 00-478 Warszawa, Poland}
%\affil{$^{O2}$Institute of Astronomy, University of Cambridge, Madingley Road, Cambridge CB3 0HA, UK}
%\affil{$^{O3}$Universidad de Concepci\'on, Departamento de Astronomia, Casilla 160-C, Concepci\'on, Chile}

\begin{abstract}
The co-planarity of solar-system planets led Kant to suggest that
they formed from an accretion disk, and the discovery of hundreds
of such disks around young stars as well as hundreds of co-planar
planetary systems by the {\it Kepler} satellite demonstrate that
this formation mechanism is extremely widespread.  Many moons
in the solar system, such as the Galilean moons of Jupiter,
also formed out of the accretion disks that coalesced into
the giant planets.  We report here the discovery of an intermediate
system OGLE-2013-BLG-0723LB/Bb composed of a Venus-mass planet 
orbiting a brown dwarf, which may be viewed either as a scaled 
down version of a planet plus star or as a scaled up version of 
a moon plus planet orbiting a star.  The latter analogy can be 
further extended since they orbit in the potential of a larger, 
stellar body.  For ice-rock companions formed in the outer parts 
of accretion disks, like Uranus and Callisto, the scaled masses
and separations of the three types of systems are similar, leading
us to suggest that formation processes of companions within accretion 
disks around stars, brown dwarfs, and planets are similar.
\end{abstract}


\section{Introduction}

Brown dwarfs are intermediate in mass between stars and planets, 
making them important laboratories for testing theories about both
classes of objects, including theories about atmospheres, binarity, 
and formation mechanism.  Brown dwarfs are also potentially a 
laboratory to test theories of planet formation, but detecting 
planets that orbit brown dwarfs is challenging.  Four super-Jupiter 
planets orbiting brown dwarfs have been detected to date by three 
techniques: direct-imaging \citep{chauvin04,todorov10}, 
radial-velocity \citep{joergens07}, and microlensing \citep{han13}.
In the first three cases, the super-Jupiter planet was within
a decade in mass of its host and well separated from it, suggesting 
that these systems more resemble scaled down binary stars than 
they do the extreme mass-ratio systems that are likely to emerge 
from accretion disks.  In the fourth case, the separation was 
substantially closer and the mass ratio somewhat more pronounced 
so that the planet could have formed through the core-accretion 
scenario that is thought to be the origin of the solar system's 
gas giants.

With present technology, substantially lower-mass companions to 
brown dwarfs, composed primarily of ice and rock, can only be 
discovered using the gravitational microlensing technique.  
Microlensing does not rely on light from either the host or planet,
but rather infers the existence and properties of these bodies
from their deflection and magnification of light from a more
distant star that is fortuitously aligned with the system.  As
such, the low-luminosity of the brown dwarf and almost complete 
absence of light from its low-mass companion do not interfere in 
any way with the discovery process.

In this paper, we report the microlensing discovery of an intermediate
system OGLE-2013-BLG-0723LB/Bb composed of a Venus-mass planet 
orbiting a brown dwarf.
The system may be viewed either as a scaled 
down version of a planet plus star or as a scaled up version of 
a moon plus planet orbiting a star, 
suggesting that formation processes of companions within accretion 
disks around stars, brown dwarfs, and planets may be similar.

% Figure 1 ----------------------------------------------------
%\begin{figure*}[th]
\begin{figure*}[th]
\epsscale{0.8}
\plotone{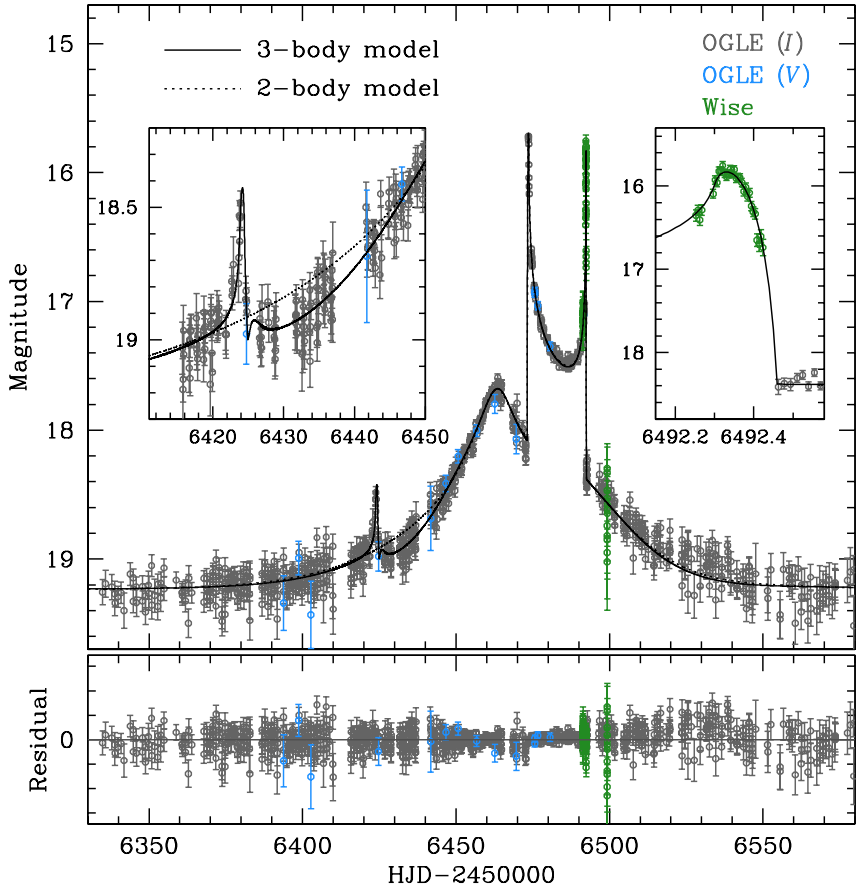}
\caption{\label{fig:one}
Light curve of the microlensing event OGLE-2013-BLG-0723,
including models with and without the planet.
Left inset shows the planetary anomaly, which includes not just the
obvious spike, but also a more extended low level depression.
Right inset shows dense coverage of caustic exit, which permitted
measurement of the Einstein radius $\theta_{\rm E}$.
}
\end{figure*}
% -------------------------------------------------------------

\section{Observation}

\begin{figure}[th]
\epsscale{1.15}
\plotone{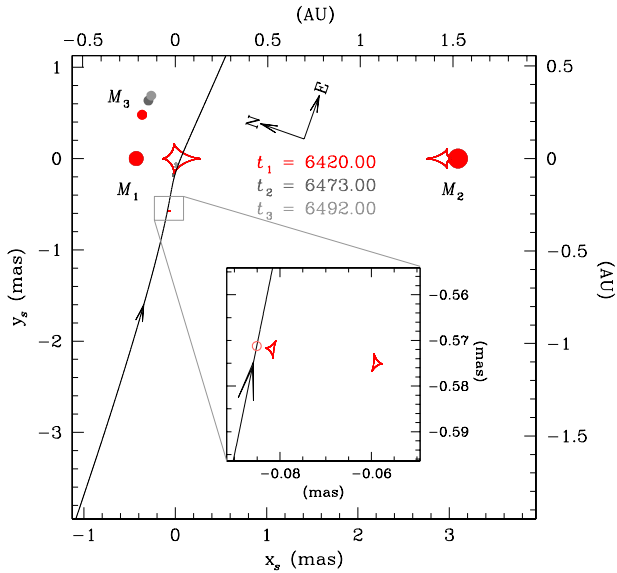}
\caption{\label{fig:two}
Geometry of the lensing event. The curve 
with an arrow represents the source trajectory. Locations of 
lens components are marked by filled dots, where $M_3$ is the 
planet, $M_1$ is the planet's host, and $M_2$ is the binary 
companion. The cuspy closed curves are caustics formed by the 
lens.  The inset shows an enlarged view around the caustic 
that produced the short-term anomaly.  The small empty circle on the 
source trajectory in the inset represents the source size relative 
to the caustic.  We note that the position of the planet and the 
resulting caustic vary in time due to its orbital motion around 
its host.   Lower/left axes are expressed by angle on the sky in mas, while 
upper/right axes show projected size of the system in AU.  
The directions North and East on the sky are indicated
}
\end{figure}

Microlensing event OGLE-2013-BLG-0723 was discovered by the Optical 
Gravitational Lensing Experiment (OGLE-IV) \citep{udalski15} on May 
12, 2013 (${\rm HJD}\sim2456424.5$) using its 1.3m Warsaw telescope 
at the Las Campanas Observatory in Chile.  
At $(\alpha,\delta)_{\rm J2000}$=$(17^{\rm h}34^{\rm m}40^{\rm s}\hskip-2pt.51$,
$-27^{\circ}26^{'}53^{''}\hskip-2pt .2)$ [$(l,b)=(-0^{\circ}
\hskip-2pt .02,2^{\circ}\hskip-2pt .83)$], 
the event lies in one of nine fields toward the Galactic bulge that 
OGLE observes $\gtrsim 1\,{\rm hr}^{-1}$, which is sufficient cadence 
to discover terrestrial planets.  In fact, the discovery was triggered 
by a short $\sim 1\,$day, $\sim 40\%$ brightening of the lensing 
event that was still in progress at the time of the announcement 
(Figure \ref{fig:one}, left inset), and which was soon suggested 
to be of planetary origin.  Hence, when the event underwent a second 
much larger (factor 10) anomalous brightening 49 days later, OGLE 
immediately recognized that it would be crucial to fully characterize 
this second anomaly.  In particular, since such violent brightening 
is always caused by the source entering a ``caustic'' (closed curve 
of formally infinite magnification -- see Figure~\ref{fig:two}), a 
caustic exit was inevitable.  Because these caustic crossings typically 
last only an hour or so and can occur at any time during the day, OGLE 
contacted the Microlensing Follow Up Network ($\mu$FUN) which operates 
a global network of narrow-angle telescopes capable of near 24-hour 
monitoring of individually important events.  With the aid of 
real-time modeling that accurately predicted the time of the exit, 
the Wise Observatory 0.46m telescope at Mitzpe Ramon, Israel, obtained 
dense coverage of this exit (Figure~\ref{fig:one}, right inset).

The principal data were taken in $I$ band.  In addition, OGLE obtained
some $V$-band data in order to characterize the microlensed source.  
Finally, $\mu$FUN obtained $H$-band observations from the 1.3m SMARTS 
telescope at Cerro Tololo Interamerican Observatory in Chile.  These 
are not used in the present analysis but will be important for the 
interpretation of future follow-up observations to cross-check the 
lens system parameters

The light curve of OGLE-2013-BLG-0723 shows a systematic decline 
in the baseline. %See the upper panel of Figure S2. 
A similar long term linear trend (of opposite sign)
was seen in OGLE-2013-BLG-0341 and was eventually traced 
to a nearby bright star that was gradually moving toward (in that
case) the source star, so that more of its flux was being ``captured''
in the tapered aperture used to estimate the source flux.
We searched for such a moving bright star by examining the
difference of two images, from 2004 and 2012.  We indeed find
a dipole from a bright star roughly $1.5^{\prime\prime}$ away, which
is the characteristic signature of such moving stars.  Having
identified the cause of this trend, we fit for it and remove it.

Photometry of the event was done using a customized pipeline 
based on the Difference Imaging Analysis method
\citep{alard98, wozniak00}. Because the photometric errors estimated by 
an automatic pipeline, $e_0$, are often underestimated, the error bars 
are readjusted by 
\begin{equation}
e=k(e_0^2+e_{\rm min}^2)^{1/2}, 
\label{eq1}
\end{equation}
where $e_{\rm min}$ 
is a term used to make the cumulative distribution function of $\chi^2$ 
as a function of lensing magnification becomes linear and the other term 
$k$ is a scaling factor used to make $\chi^2$ per degree of freedom (dof) 
becomes unity. The former process is needed in order to ensure that the
dispersion of data points is consistent with error bars of the source
brightness and the latter process is required to ensure that each data
is fairly weighted according to error bars.

% Figure 2 ----------------------------------------------------
% -------------------------------------------------------------

\section{Light-curve Analysis}

The principal features of the light curve are a large double-horned
profile near the peak, which is a signature of binary microlenses,
and a short-lived spike two months earlier, which is characteristic
of planetary companions.  Modeling these features yields the mass
ratios and instantaneous locations of the three bodies relative to
the source trajectory.  See Figure~\ref{fig:two}.  However, by 
themselves, these features do not yield either the physical or angular 
scale of the diagram, nor its orientation on the sky.  Rather, the diagram
is scaled to the ``Einstein radius'', which can be expressed either
as an angle $\theta_{\rm E}$, or as a physical length $\tilde r_{\rm E}$ projected
on the observer plane.

The source star and the Earth's orbit act as ``rulers'' to determine
these two quantities, as well as the angular orientation.  The source 
angular radius $\theta_*$ is known from its observed temperature and 
flux.  Hence, when it passes over a caustic (see Figure~\ref{fig:two}),
the finite extent of the source smears out what would be an extremely 
sharp peak for a point source (Figure~\ref{fig:one}, right inset).
Second, Earth's accelerated motion induces distortions in the light curve, 
which fix both $\tilde r_{\rm E}$ relative the size of Earth's orbit (1 AU) 
and the orientation of the diagram relative to that of Earth's orbit, namely 
the ecliptic.  Hence, we are able to display both the angular scale and 
physical orientation in Figure 2.  The mass $M$ and distance $D_L$ of the 
lens are then given by
\begin{equation}
{1\over D_L} = {\theta_{\rm E}\over \tilde r_{\rm E}} + {1\over D_S},
\qquad
{4G M\over c^2} = \tilde r_{\rm E} \theta_{\rm E} .
\label{eq2}
\end{equation}
The first equation follows from simple geometry.  Once the distance
is known, the second equation follow from Einstein bending angle
$\alpha = 4 G M/(b c^2)$, where $b$ is the impact parameter of the 
lens-source approach.  See the graphical presentation of the formalism 
in \cite{gould00}.

% Table 1 =============================================
\begin{deluxetable}{ll}[lr]
\tablecaption{Lensing parameters\label{table:one}}
\tablewidth{0pt}
\tablehead{
\multicolumn{1}{c}{Quantity} &
\multicolumn{1}{c}{Value}
}
\startdata
$\chi^2/{\rm dof}$            &  4126.8/4073 \\
$t_0$ (HJD')                  &  6484.526 $\pm$ 0.037 \\
$u_0$                         & -0.079    $\pm$ 0.002 \\
$t_{\rm E}$ (days)            &  68.48    $\pm$ 0.01 \\
$s_1$                         &  5.07     $\pm$ 0.02 \\
$q_1$                         &  3.11     $\pm$ 0.02 \\
$\alpha$ (rad)                & -1.195    $\pm$ 0.003 \\
$s_2$                         &  0.97     $\pm$ 0.02 \\
$q_2$ $(10^{-5})$             &  6.61     $\pm$ 0.01 \\
$\psi_0$ (rad)                & -4.936    $\pm$ 0.005 \\
$\rho$ $(10^{-3})$            &  1.40     $\pm$ 0.02 \\
$\pi_{{\rm E},N}$             & -0.05     $\pm$ 0.01 \\
$\pi_{{\rm E},E}$             &  1.35     $\pm$ 0.02 \\
$ds_2/dt$ $({\rm yr}^{-1})$   &  0.81     $\pm$ 0.02 \\ 
$d\psi/dt$ $({\rm yr}^{-1})$  & -0.50     $\pm$ 0.02  \\
$I_{\rm S}$                   &  19.825   $\pm$ 0.006 \\
$I_{\rm B}$                   &  20.181   $\pm$ 0.008
\enddata  
\tablecomments{ 
$I_{\rm S}$ and $I_{\rm B}$ represent the baseline magnitudes of the 
source and blend, respectively.
}
\end{deluxetable}
% ------------------------------------------------------------

% Figure 3 =============================================
\begin{figure*}[th]
\epsscale{0.8}
\plotone{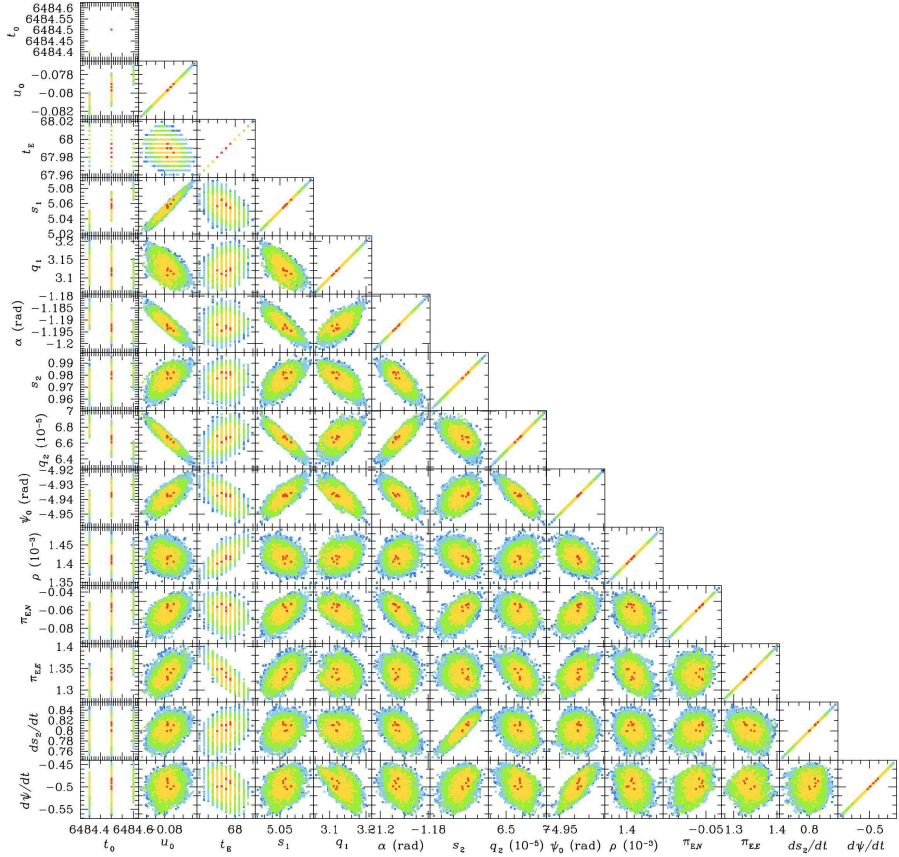}
\caption{\label{fig:three}
Distribution of the microlensing parameters of the best-fit solution,
whose central values and errors are shown in Table S1. Color coding 
indicates points on the Markov Chain within 1 (red), 2 (yellow), 
3 (green), 4 (cyan), 5 (blue) $\sigma$ of the best fit.
}
\end{figure*}
% -------------------------------------------------------------

\subsection{Microlens Modeling}

Full modeling of the light curve ultimately requires a total
of 19 parameters, including 14 parameters to describe the lensing 
system and five flux parameters.  Some of these manifest themselves 
in quite subtle effects.  The modeling therefore proceeded in several 
distinct phases.  %First, an underlying linear trend in the OGLE baseline 
%data was identified and removed.  See Section~\ref{sec:trend}
First, we determine the principal parameters describing the binary-lens system
by removing the data for 25 days around the planetary
perturbation and fitting the remaining light curve to seven system 
parameters plus four flux parameters (two for each observatory).  
These nine include three parameters to describe the underlying point-lens 
event $(t_0,u_0,t_{\rm E})$, three to describe the perturbations induced by 
the binary companion $(s_1,q_1,\alpha)$, one to describe the ratio of 
the source radius to the Einstein radius, $\rho=\theta_*/\theta_{\rm E}$,  
and the two-dimensional microlens parallax vector 
$\pivec_{\rm E}=(\pi_{{\rm E},N},\pi_{{\rm E},E})$.

Here $t_0$ is the time of closest approach to a reference point in
the geometry, which is chosen as the ``center of magnification'' of 
the caustic traversed by the source, $u_0$ is the impact parameter
in units of $\theta_{\rm E}$, and $t_{\rm E}$ is the time required for the source
to cross $\theta_{\rm E}$.  We choose $\theta_{\rm E}$ as the Einstein radius
of the mass associated with the caustic, which is the total mass for 
the close-binary case, and the mass of the binary component located 
closer to the source trajectory for the wide-binary case.  This convention 
leads to similar $t_{\rm E}$ for the two cases, e.g.\ \citep{gould14}.

The three binary parameters are the separation of the components in
units of $\theta_{\rm E}$, the ratio of the companion to primary component
masses, and the angle of the source trajectory relative to the binary
axis at $t_0$ (since parallax causes this to change with time).

The amplitude of the parallax vector is $\pi_{\rm E}={\rm AU}/\tilde r_{\rm E}$, 
while its direction is that of the lens-source relative proper motion (in the 
geocentric frame).  In contrast to the lens-source trigonometric relative 
parallax $\pi_{\rm rel}=\pi_{\rm E}\theta_{\rm E}$, which is a scalar quantity, 
the microlens parallax requires two parameters because the lens is not directly 
observed, so the orientation of the lens-source separation relative to the 
ecliptic is not known a priori. See \citet{gouldhorne}.  Normally, one would
not include parallax in the initial fit, but in this case it is so large that 
the projected Einstein radius is substantially smaller than Earth's orbit, 
$\tilde r_{\rm E}\sim 0.35\,{\rm AU}$, so that there is no even approximately 
good solution without including $\pivec_{\rm E}$ in the fit.

With this parameterization, we find an excellent fit for the wide-binary
case $(s>1)$ but only a crude approximation to the observed light curve
for $s<1$.  We then incorporate two-parameter orbital motion, essentially
the instantaneous relative transverse velocity of the components
in Einstein-radius units.  This refinement results in negligible
improvement for the $s>1$ solution, but dramatic improvement in
the $s<1$ solution, so that the two models provide qualitatively similar
approximations of the data.  However, it is found that the $s<1$ case is 
unlikely due to various reasons as discussed in Appendix.

We then restore the data from the planetary perturbation and add
five additional parameters to the fit: $(s_2,q_2,\psi_0,\gammavec)$,
where $\gammavec = (ds_2/dt/s_2,d\psi/dt)$.  Here $(s_2,q_2)$ are
the separation and mass ratio of the planet relative to its host
(the smaller component of the binary), $\psi_0$ is the angle between
binary axis and the planet-host axis at $t_0$, and $\gammavec$ is the
two dimensional projected velocity of the planet relative to its host.
Table~\ref{table:one} shows the best fit parameters.  In Figure~\ref{fig:three}, 
we also present the distribution of $\Delta\chi^2$ in the parameter 
space in order to show the uncertainties and correlations between the 
parameters.

% Figure 4 =============================================
\begin{figure}[th]
\epsscale{1.15}
\plotone{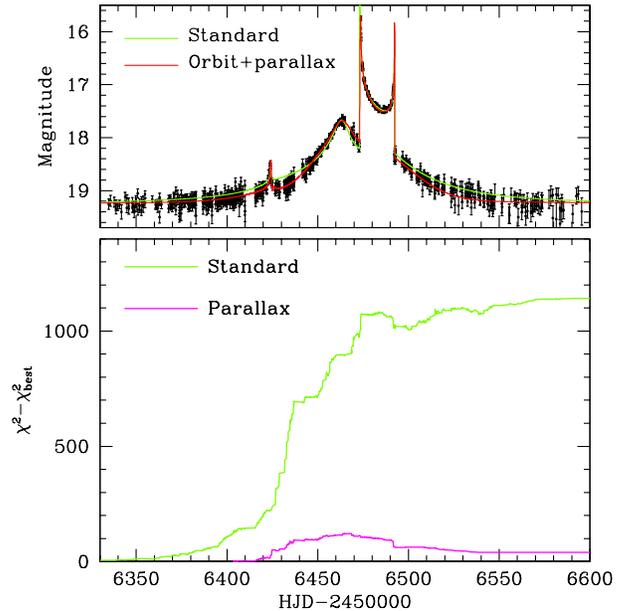}
\caption{\label{fig:four}
Cumulative distributions of $\Delta\chi^2$ from the best-fit 
(orbit+parallax) model for the standard (green curve) and parallax 
(magenta curve) models. 
}
\end{figure}
% -------------------------------------------------------------

\subsection{Origin of Parallax and Orbital Signals}

The microlens parallax signal is quite strong ($\Delta\chi^2>1100$),
but since the amplitude $\pi_{\rm E}=2.7$ is one of the largest ever
detected and hence quite unusual. Therefore, it is prudent to check 
whether this signal is coming mainly from  one part of the light 
curve, which would be suspicious, or from the entire magnified
portion of the light curve, as one would expect.

Figure~\ref{fig:four} shows that in fact the signal is coming from the
whole magnified light curve.  It also shows that planetary orbital 
signal, while much weaker ($\Delta\chi^2=41$), is also broadly based.  
As expected, the orbital signal is more compact than the parallax
signal.

\subsection{Degeneracy}

Because the light curve is well covered, the parameters are well 
determined.  However, the solution is still subject to three well 
known discrete degeneracies.  First, there is a degeneracy between 
pairs of companion mass ratio and normalized separation $(q,s)$, 
with one lying inside the Einstein radius $(s<1)$ and the other outside.
This degeneracy is fully understood at first \citep{griest98,dominik99} 
and second \citep{an05} order for the case that the binary is approximated 
as static.  We find that within this context there is no ``close'' 
solution, i.e., $\Delta\chi\sim 2000$ between close and wide solutions.  
However, if both close and wide binary models are permitted two degrees 
of orbital motion, i.e. $ds_2/dt$ and $d\psi/dt$, then the wide solution 
shows essentially no improvement while the close solution improves 
dramatically and matches the wide solution.  In Appendix, we discuss 
the mathematical and physical nature of this new variant of the close/wide 
degeneracy and show that it implies that the wide solution is strongly 
favored.  We also present additional corroborating arguments that this 
is the case.  We therefore consider only the wide solution.

Second, if the source lay on the ecliptic, the entire solution could 
be ``flipped'' relative to the ecliptic without changing the light 
curve.  Since the source lies just $4^\circ$ from the ecliptic, this 
degeneracy is close to exact, with the preferred solution favored by 
only $\Delta\chi^2 = 38$.  While this is enough to clearly favor one
solution, the main point is that the two solutions yield almost identical 
physical implications.

Finally, Figure~\ref{fig:two} shows the 
small planetary caustics just to the right of the source trajectory.
In principle, there could be another solution with these caustics 
just to the left.  However, this geometry is unable to reproduce the
long slow ``depression'' that follows the planetary anomaly and is
due to a demagnification ``valley'' that follows the axis defined
by the two small caustics.  Hence, the solution is both unique and
measured with good precision.

% Figure 5 =============================================
\begin{figure}[t]
\epsscale{1.15}
\plotone{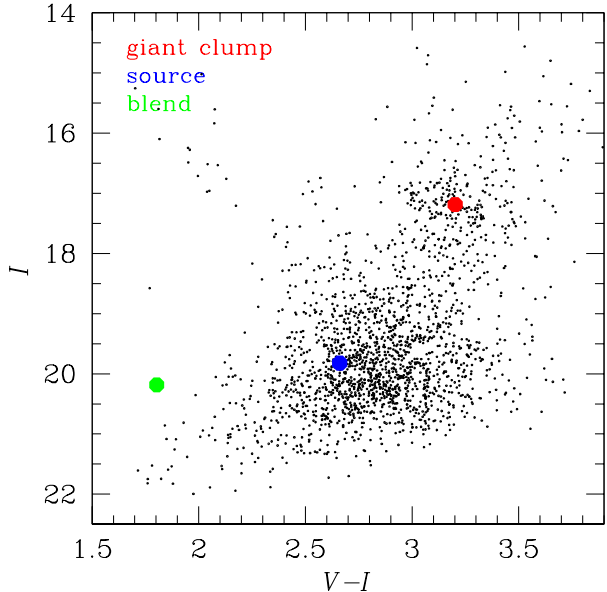}
\caption{\label{fig:five}
Location of the source star in the color-magnitude 
of stars in the nearby field. Also marked are the positions of the 
centroid of the red clump giant stars and of the blend.  For the
first two, the error bars are smaller than the points.  For the
last, the magnitude is secure but the error in the color (not shown)
is both large and uncertain, being at least 0.2 mag.
}
\end{figure}
% -------------------------------------------------------------

\subsection{Characteristics of the Source Star and Blended Light}

We determine the angular source radius $\theta_*$ based on its 
de-reddened color $(V-I)_0$ and brightness $I_0$ \citep{ob03262},
which we determine from their offsets from the centroid of 
the red giant clump $[(V-I),I]_{0,\rm clump} = (1.06,14.45)$ 
\citep{bensby11,nataf13},
under the assumption that the source and the clump giants
suffer similar extinction.  
Figure~\ref{fig:five} gives a graphic representation of this
procedure, and also shows the position of the star (or ensemble
of stars) that are blended with the source.  That is, in each band,
the light-curve flux is modeled as $F(t) = f_s A(t) + f_b$, where $A$
is the magnification.  The source and blend magnitudes are then
derived from the fit parameters $f_s$ and $f_b$, respectively.

We find $(V-I,I)_0= (0.52,17.09)$. From this color and flux,
we derive $\theta_*=0.98\pm 0.08\ \mu{\rm as}$ by converting $(V-I)\rightarrow (V-K)$
\citep{bessell88} and then applying a color/surface brightness relation 
\citep{kervella04}.  The derived color is well blueward of
the overwhelming majority of Galactic bulge stars, except for
horizontal branch stars, which are substantially brighter.  Hence,
the source is almost certainly in the Galactic disk.  Disk sources
are rare in general (a few percent) partly because the source must
be farther than the lens with the lenses being mostly in the bulge
or fairly far in the disk, and partly because the column density of
bulge stars is higher than for disk stars.  However, given that
this particular lens lies far in the foreground $(D_L\sim 0.5\ {\rm kpc}$),
it is not unexpected for the source to lie in the disk.

Based on the source color and magnitude, we estimate its distance
$D_S = (7\pm 1)\ {\rm kpc}$.  The exact distance plays no role in the
determination of $\theta_*$ because this depends only on the flux 
$F\propto L/D_S^2$, not the luminosity $L$ and distance separately,
provided that the source lies behind essentially all the dust.
Given that the source sits 360 pc above the Galactic plane, this is 
almost certainly the case.  The uncertainty in the source distance also
enters into the uncertainty in the lens distance through Equation~(\ref{eq2}).
In practice, however, this leads to an error in source distance of only 4\%,
which is small compared to other sources of errors.

Finally, we note that the source color indicates a temperature
$T\sim 6100\,$K, which leads to an estimate of the limb-darkening
coefficient (needed for modeling the caustic exit) of $u=0.495$
\citep{claret00}.

In most microlensing events, the blended light $f_b$ is just a nuisance
parameter, but in the present case it is potentially very important.
We first note that while the $I$-band magnitude is secure, the error
in the $(V-I)$ color (not shown) is quite large.  First, the formal
error of the fit is 0.2 mag.  In addition, in contrast to the source
flux, which is determined solely from changes in brightness, the
blend flux depends on the accuracy of the baseline photometry.  This
can be corrupted by bright nearby blends, which are plainly visible on 
the images, and by stars that are too faint to be identified.  Hence,
we consider the color to be only indicative and do not make use of it
in further discussion.  However, the $I$ band measurement is too
bright to be the OGLE-2013-BLG-0723LA by about 1.5 mag.

To further understand the origin of this light, we measure the
position of the source relative to the ``baseline object'', which
is an unresolved blend of the source with the star or stars giving
rise to $f_b$. We find $\Delta\theta_{\rm base} = 55\,$mas.  Since 
the blend contains 43\% of the baseline light, this implies that it 
is naively offset from the source by 
$\Delta\theta_{\rm blend} = \Delta\theta_{\rm base}/0.43=125\,$mas.
However, while
the source position can be measured extremely well because it
becomes highly magnified, the ``baseline object'' is subject
to contamination from unseen stars as well as the wings of nearby
bright stars.  We conclude that the centroid of blend light is
separated from the source by $\Delta\theta_{\rm blend} \lesssim 125\,$mas.
Because the surface density of stars $I<20.1$ in this
field is $\lesssim 0.1\,{\rm arcsec}^{-2}$ \citep{holtzman98},
the probability that this blended light lies so close to the source 
(and lens) by chance is $<1\%$.  It is therefore most likely a
companion to either the source or lens.  This issue can be resolved
by obtaining AO images at two epochs to determine whether its
proper motion is consistent with the source or lens.  If it proves
to be associated with the lens, then this is a four-body system
consisting of a brown dwarf orbited by a Venus-mass planet, with
both orbiting a very low-mass star, and with this entire group
orbiting another star that is roughly double the mass of the
star that contributed to the microlensing event.

% Table 2 =============================================
\begin{deluxetable}{ll}[h]
\tablecaption{Physical parameters\label{table:two}}
\tablewidth{0pt}
\tablehead{
\multicolumn{1}{c}{Quantity} &
\multicolumn{1}{c}{Value}
}
\startdata
Mass of the planet                                 & $0.69  \pm 0.06\ M_\oplus$ \\
Mass of the host                                   & $0.031 \pm 0.003\ M_\odot$ \\
Mass of the binary companion                       & $0.097 \pm 0.009\ M_\odot$ \\
Distance to the lens                               & $0.49  \pm 0.04$ kpc     \\
Projected planet-host separation                   & $0.34  \pm 0.03$ AU      \\
Projected separation between binary components     & $1.74  \pm 0.15$ AU      \\
Relative lens-source proper motion (geocentric)    & $3.75  \pm 0.33$ mas/yr \\
Relative lens-source proper motion (heliocentric)  & $14.5  \pm 1.3$ mas/yr 
\enddata  
%\tablecomments{ 
%Physical quantities.
%}
\end{deluxetable}
% ------------------------------------------------------------

\subsection{Physical Parameters}

The principal physical parameters are given in Table~\ref{table:two}.  
% -----
From the determined masses of the individual lens components, it 
is found that the lens is a triple system that is composed of a 
Venus-mass planet (0.69~$M_\oplus$) orbiting a 
brown-dwarf host (0.031~$M_\odot$) that is accompanied 
by a low-mass star (0.097 $M_\odot$). The projected planet-host 
separation is 0.34~AU, while the separation 
between the binary components is 1.7~AU.  The system 
is located at a distance 490~pc toward the Galactic center.

% -----
Note that 
we give the proper motion in both the geocentric frame in which the 
measurements were actually made, and the heliocentric frame, which 
will be useful for comparison with future observations.  Since $D_L$ 
is known, the conversion from geocentric to heliocentric frames is 
straightforward.

\subsection{Which Binary Component is Host to the Planet?}

From Figure~\ref{fig:two} or Table~\ref{table:one}, the planet's
projected position is $5.07/0.97=5.2$ times closer than the
stellar companion.  If the planet were orbiting the more massive 
companion, this would occur with probability $<4\%$.  In addition, 
in order to ensure stability of the 3-body system,  the brown dwarf 
would have to be behind (or in front of) the star-planet system by 
a at least factor of 3 larger distance than the projected separation 
of the brown dwarf and star.  This chance alignment further reduces 
the probability by a factor $3^2$.  Hence, while this configuration 
is possible in principle, it has extremely low probability.  A very 
similar argument applies to a circumbinary configuration, in which
the planet's orbit is at least three times larger than the
separation between the brown dwarf and star, but just happens
to be seen in projection close to the brown dwarf.

\section{Discussion}

OGLE-2013-BLG-0723LBb is a missing link between planets and moons.
That is, its host OGLE-2013-BLG-0723LB, being a brown dwarf in orbit 
around a self-luminous star, is intermediate between stars and planets, 
in both size and hierarchical position.  Moreover, the scaled mass and 
host-companion separation of OGLE-2013-BLG-0723LB/Bb are very similar 
to both planets and moons in the solar system.  See Table~\ref{table:three}.
Both Uranus and Callisto are believed to have formed in the cold
outer regions of their respective accretion disks, and are mostly
composed of the raw materials of such regions: ice with some rock.
In the case of Uranus, it is believed to have been formed closer to 
the current location of Saturn (10 AU) and to have migrated outward.  
In the table, the companion-host separations are scaled to the host 
mass.  This is appropriate because the ``snow line'', the inner radius 
at which icy solids can form (2.7 AU in the solar system) increases 
with host mass, probably roughly linearly.  A plausible inference from 
the first three lines of Table 2 is that these processes scale all the 
way from solar-type stars hosting planets, to brown dwarfs hosting
``moon/planets'', to giant planets hosting moons.

However, while OGLE-2013-BLG-0723LBb is almost certainly orbiting 
OGLE-2013-BLG-0723LB today, we 
cannot guarantee that it was born there.  Planets in relatively close
binaries (like OGLE-2013-BLG-0723LA,B) can jump from one star (or in 
this case, brown dwarf) to the other, if the system is dynamically perturbed 
\citep{kratter12}.  If this is what happened to OGLE-2013-BLG-0723LBb, then  
would have been captured from an orbit around OGLE-2013-BLG-0723LA to 
its current orbit around OGLE-2013-BLG-0723LB.  It would then be similar 
to Triton, which was captured by Neptune out of solar orbit.  Table~\ref{table:three} 
shows that the Neptune-Triton system in its current configuration is 
also not terribly different than a scaled version
of OGLE-2013-BLG-0723LB/Bb.

It will be interesting to model the dynamics of the triple system
OGLE-2013-BLG-0723LA,B,Bb, but it would be premature to do so.  This 
is because there is likely a fourth member of this system that is 
more massive and luminous than the other three, and which is separated by
of order 100 AU or less.  If confirmed, this member could significantly
impact the system dynamics via the Lidov-Kozai mechanism.  The
principal evidence for this fourth member is that there is excess light
in the aperture that is too bright to be the primary star
OGLE-2013-BLG-0723LA, and is so closely aligned with the source that
it is unlikely to be a random interloper.  Because the source flux
is known from the $H$-band light curve (which is not included in the 
present analysis), it will be straightforward to subtract out the source 
light from a high-resolution adaptive optics (AO) image and determine the 
exact location of the blended light.  If well separated, then its association
with the lens can be established from its proper motion, derived from 
observations at two epochs, which can be compared with Table~\ref{table:two}.
If directly superposed in the AO images, then it is almost
certainly associated with either the source or the lens because of
the extremely low probability of a random interloper.  The two possibilities
can be distinguished based on the apparent color because the source
is highly reddened by dust, but the lens sits in front of the great
majority of the dust.

OGLE-2013-BLG-0723LBb is the second terrestrial mass planet orbiting
one member of a binary system composed of two very low mass
objects.  The other was OGLE-2013-BLG-0341LBb.  Gould et al. \citep{gould14}
showed that if all low-mass stars were in such systems, roughly one should 
have been detected.  They noted that no strong statistical 
inference could be derived from this fact because, for example, if only 
1\% of low-mass stars were in such systems, there would be a 1\% chance 
to find one.  Because the question about the frequency of these systems 
was being framed a posteriori, a 1\% effect is not particularly significant.

However, OGLE-2013-BLG-0723LBb reproduces three of the four elements 
that went into their estimate of the frequency of such detections, namely 
small probability of passing by the planetary caustic, small probability 
of crossing the central (binary) caustic, and location in a high-cadence 
OGLE fields.   The fourth criterion, source magnitude brighter than 
$I_S=18.5$, is not satisfied, since $I_S\sim 19.8$.  If the limit were 
extended to $I_S<20$, the total number of available events would roughly 
double.  Hence, the prediction would rise to two.   With two predicted
and two detected, the probability that these are extremely rare drops 
roughly as the square, i.e., roughly $10^{-4}$ probability of this occurring 
by chance if only 1\% of these stars were in such systems.  Hence, while 
there is still no compelling evidence that these systems are extremely 
common, at least they are not extremely rare.  Further, the current 
detections remain consistent with the hypothesis that these systems are 
ubiquitous.

% Table 2 ----------------------------------------------------
\begin{deluxetable}{lcc}[h]
\tablecaption{Companion/Host pairs\label{table:three}}
\tablewidth{0pt}
\tablehead{
\multicolumn{1}{c}{Host/Companion} &
\multicolumn{1}{c}{$M_{\rm comp}/M_{\rm host}$} &
\multicolumn{1}{c}{$r_{\rm c-h}/M_{\rm host}$} \\
\multicolumn{1}{c}{} &
\multicolumn{1}{c}{($10^{-5}$)} &
\multicolumn{1}{c}{(${\rm AU}/M_\odot$)} 
}
\startdata
Uranus/Sun                &     4.4 &    19 \\
OGLE-2013-BLG-0723LBb/B   &     6.6 &    11 \\
Callisto/Jupiter          &     5.7 &    13 \\
\hline
Triton/Neptune            &    21   &    46 
\enddata  
%\tablecomments{ 
%Physical quantities.
%}
\end{deluxetable}
% ------------------------------------------------------------

\acknowledgments

The OGLE project has received funding from the National Science Centre,
Poland, grant MAESTRO 2014/14/A/ST9/00121 to AU.
Work by C.H. was supported by Creative Research Initiative Program
(2009-0081561) of National Research Foundation of Korea. 
A.G. and B.S.G. acknowledge support from NSF AST-1103471. 
B.S.G., A.G., and R.W.P. acknowledge support from NASA grant NNX12AB99G.

\appendix
%\begin{center}
%APPENDIX\\
%\smallskip
%BINARY-LENSING PARAMETERS
%\end{center}
\section{RESOLUTION OF THE CLOSE/WIDE DEGENERACY}

\smallskip 

In the main text of the paper, we stated that the wide solution 
was consistent with a static binary while the close solution required 
substantial orbital motion to match the wide-solution light curve.  
At first sight this seems qualitatively reasonable since close binaries 
have shorter periods than wide ones.  However, only one of these solutions 
corresponds to the physical system, while the other is merely a mathematical
reflection of it.  Hence, the real question is not whether each solution 
is physically plausible but what is the relative likelihood where each 
solution, if it were real, could generate the other as a doppelganger.

Before addressing this, some comments must be in order as why this 
degeneracy is appearing in this case and how likely it will be to
appear generically.  In the simplest cases 
two solutions appear because to first order, the quadrupole expansion
of the close-binary potential and the tidal expansion of the wide-binary
potential are identical and so (to this order) produce identical
caustics \citep{griest98,dominik99}.  
Contemporaneously, it was noticed that in more typical
cases, the degeneracy could remain extremely severe even though
the caustics were not actually identical and, indeed, appeared rotated
relative to each other (Fig. 8 of \citep{afonso00}).  This was ultimately
explained by carrying out the analysis to second order \citep{an05}.
That is, an additional degree of freedom (basically, rotation) could
compensate for the slower convergence of the potential.  
For wide binaries, the potential converges more slowly as the mass ratio
of the companion increases and as its separation gets closer.  This is
well illustrated by small mass-ratio, e.g., planet/star, systems.
Topologically, the two caustics are the same, but geometrically they
are very different.  The planet has a relatively weak effect on the
stellar potential, meaning that the star's caustic (``central caustic'') 
is well-described by low-order terms and hence subject to the close/wide
degeneracy.  By contrasts, the planetary caustic is strongly perturbed
and not subject to this degeneracy (unless the separation is very large).

In the present case, the caustic feature in the light curve of the wide 
solution is associated with the lower-mass companion, and the mass ratio 
is relatively large.  Hence, one expects that (as in the planetary case) 
higher order terms will play a relatively large role.  And indeed they
prove to be too large to permit the degeneracy analyzed by \citep{an05}.
Orbital motion then provides the additional degree of freedom to 
compensate for the non-trivial higher-order terms.

Consider then a wide binary that is qualitatively similar to the present 
case.  Regardless of details, its light curve can always be mimicked 
by including the appropriate amount of orbital motion.  On the
other hand, consider a (real) close binary with the $(s,q)$ parameters
predicted by the wide binary but with so far unspecified orbital
parameters.  Can this system ``generate'' a static wide binary
with the same light curve?  The answer is: only if its orbital
parameters happen to lie in a narrow range around those predicted by the
wide pair.  Otherwise, the static wide binary would have been consistent 
with a wide range of close-binary orbital motions.  Because $\Delta\chi^2$
is quite large, the fine-tuning requirements of the close solution are
severe.  This is the principal reason that we discount it.

However, in addition, if the close solution is spurious and merely
tuned to the mathematical requirements of matching the wide
solution, it is likely to have characteristics that are atypical
of real systems, which can be a further indication of its spurious
nature.  We note two such features.  First, the ratio of projected
kinetic to potential energy \citep{dong09} is quite low
\begin{equation}
\beta\equiv \biggl({E_{\rm kin}\over E_{\rm pot}}\biggr)_\perp = 
{(s\theta_{\rm E} D_L/{\rm AU})^3(\gamma\,{\rm yr})^2\over M/M_\odot} = 0.023
\label{eqn:beta}
\end{equation}
compared to typical values $0.1\lesssim \beta\lesssim 0.5$.  The three physical
effects that can produce low $\beta$ are 1) it is near apocenter
of an extremely eccentric orbit, 2) our line of sight happens to
be almost tangent to the orbital velocity vector, or 3) we happen
to view the system edge on, so that its semi-major axis is much
larger than its projected separation.  Since the unprojected
ratio typically has values $E_{\rm kin}/E_{\rm pot}\sim 0.5$
reducing this number by a factor $\sim 20$ via projection effects
requires some fine tuning.  However, in the present case there
is an additional complication: the third option generates the
need for an additional low probability event.  That is, if the
two stars are extended along the line of sight, then either
this projection must be very great to allow the planet to have a stable
orbit around one star (actually brown dwarf), to the planet
itself must be seen in projection next to the pair despite the
fact that its orbit is at least three times that of the extended
binary.  Finally, we note that the source star is four times fainter
in the close model.  Since its color (so spectral type) is still the
same, this places it twice as far as in the wide solution, where
the density of such stars is about a factor 40 smaller.  This is
somewhat compensated by the eight times larger phase space, but
still does further reduce the probability of this solution by a
factor 5.

All the arguments given above make the close-binary solution
very unlikely but they do not exclude it.  However, this solution
can also be tested by direct imaging.  Since there are compelling
reasons to conduct such imaging, possibly at several epochs, 
no special effort is required to
do this.  If these observations reveal a blend that is substantially
brighter than the source, then the close solution will be confirmed.
The expected baseline magnitudes of the source and blend are
$(I_{\rm S},I_{\rm B})=(21.28,19.42)$, respectively. 
If the wide solution is correct, we expect the blend and source to
be comparable (depending on their $I-H$ colors).


\begin{thebibliography}{99}
\bibitem[Afonso et al.(2000)]{afonso00} Afonso, C., et al. 2000, \apj, 532, 340 
\bibitem[Alard \& Lupton(1998)]{alard98} Alard, C. \& Lupton, R.~H. 1988, \apj, 503, 325 
\bibitem[An(2005)]{an05} An, J.~H. 2005, \mnras, 356, 409
\bibitem[Bensby et al.(2011)]{bensby11} Bensby, T., et al., 2011, \aap, 533, 134 
\bibitem[Bessell \& Brett(1988)]{bessell88} Bessell, M.~S.\& Brett, J.~M. 1988, \pasp, 100, 1134 
\bibitem[Chauvin et al.(2004)]{chauvin04} Chauvin, G., et al. 2004, \aap, 425, L2 
\bibitem[Claret(2000)]{claret00} Claret, A. 2000, \aap, 363, 1081 
\bibitem[Dominik(1999)]{dominik99} Dominik, M. 1999, \aap, 349, 108
\bibitem[Dong et al.(2009)]{dong09} Dong, S., et al. 2009, \apj, 695, 442 
\bibitem[Gould(2000)]{gould00} Gould, A. 2000, \apj, 542, 785
\bibitem[Gould et al.(2014)]{gould14} Gould, A., et al., 2014, Science, 345, 46
\bibitem[Gould \& Horne(2013)]{gouldhorne} Gould, A., \& Horne, K. 2013, \apj, 779, 28 
\bibitem[Griest \& Safizedah(1998)]{griest98} Griest, K., \& Safizedah, N. 1998, \apj, 500, 37
\bibitem[Han et al.(2013)]{han13} Han, C., et al., \apj, 2013, 778, 38 
\bibitem[Holtzman et al.(1998)]{holtzman98} Holtzman, J.~A., et al. 1998, \aj, 115, 1946 
\bibitem[Joergens \& M{\"u}ller(2007)]{joergens07} Joergens, V., \& M{\"u}ller, A. 2007, \apj, 666, L113
\bibitem[Kervella et al.(2004)]{kervella04} Kervella, P., et al. 2004, \aap, 426, 297 
\bibitem[Kratter \& Perets(2012)]{kratter12} Kratter, K.~M. \& Perets, H.~B. 2012, \apj, 753, 91 
\bibitem[Nataf et al.(2013)]{nataf13} Nataf,D.~H., et al., 2013, \apj, 769, 88
\bibitem[Todorov et al.(2010)]{todorov10} Todorov, K., et al. 2010, \apj, 714, L84 
\bibitem[Udalski et al.(2015)]{udalski15} Udalski, A., Szyma{\'n}ski, M.K. \& Szyma{\'n}ski, G. 2015, Acta Astron., 65, 1
\bibitem[Wo{\'z}niak(2000)]{wozniak00} Wo{\'z}niak, P.~R. 2000, Acta Astron., 50, 421 
\bibitem[Yoo et al.(2004)]{ob03262} Yoo, J., et al., 2004, \apj, 603, 139 
\end{thebibliography}
\end{document}